\documentclass[12pt]{article}
\usepackage{amsmath}
\usepackage{amssymb}
\usepackage{latexsym}
\usepackage{amsfonts}
\usepackage{graphicx}
\usepackage{varioref}
\usepackage{ifthen}
\usepackage{amsmath,color,soul}
\usepackage{amssymb}
\usepackage[font={small}]{caption}
\begin{document}
\parindent 0mm 
\setlength{\parskip}{\baselineskip} 
\thispagestyle{empty}
\pagenumbering{arabic} 
\setcounter{page}{0}

\noindent
\begin{center}
{\large {\bf Up- and down-quark masses from QCD sum rules}}
\end{center}

\begin{center}
{\bf  C. A. Dominguez}$^{a}$, 
{\bf A. Mes}$^{a}$,
{\bf and} 
{\bf K. Schilcher}$^{a,b}$
\end{center}

\begin{center}
{\it $^{(a)}$Centre for Theoretical and Mathematical Physics and Department of Physics, University of
Cape Town, Rondebosch 7700, South Africa\\
\vspace{.3cm}
$^{(b)}$ Institut f\"{u}r Physik, Johannes Gutenberg-Universit\"{a}t\\
Staudingerweg 7, D-55099 Mainz, Germany}
\end{center}
\begin{center}
\textbf{Abstract}
\end{center}

\noindent
The QCD up- and down-quark masses are determined from an optimized QCD Finite Energy Sum Rule (FESR) involving the correlator of axial-vector current divergences. In the QCD sector this correlator is known to five loop order in perturbative QCD (PQCD), together with non-perturbative corrections from the quark and gluon condensates. This FESR is designed to reduce considerably the systematic uncertainties arising from the  hadronic spectral function. The determination is done in the framework of both fixed order and contour improved perturbation theory. Results from the latter, involving far less systematic uncertainties, are:  $\bar{m}_u (2\, \mbox{GeV}) = (2.6 \, \pm \, 0.4) \, {\mbox{MeV}}$, $\bar{m}_d (2\, \mbox{GeV}) = (5.3 \, \pm \, 0.4) \, {\mbox{MeV}}$, and the sum $\bar{m}_{ud} \equiv (\bar{m}_u \, + \, \bar{m}_d)/2$, is
$\bar{m}_{ud}({ 2 \,\mbox{GeV}})  =( 3.9 \, \pm \, 0.3 \,) {\mbox{MeV}}$. 
\newpage

\section{Introduction}
\noindent

Quark masses together with the strong coupling are the fundamental parameters of Quantum Chromodynamics (QCD). Their values at some given scale can be determined numerically from Lattice QCD (LQCD), as well as analytically from QCD sum rules (QCDSR) \cite{FLAG}-\cite{QCDSR2}. Historically, QCDSR were first formulated in the framework of Laplace transforms \cite{QCDSR1}-\cite{OLD}. As precision determinations became necessary, in order to compare results with those from LQCD, current QCDSR are formulated in the complex squared energy, s-plane, as first proposed in \cite{Shankar}. In this plane the only singularities in current correlators are along the real positive axis. They correspond to hadronic bound states on this axis, as well as resonances in the second Riemann sheet.

\begin{figure}
[ht]
\begin{center}
\includegraphics[height=2.5in, width=2.5in]
{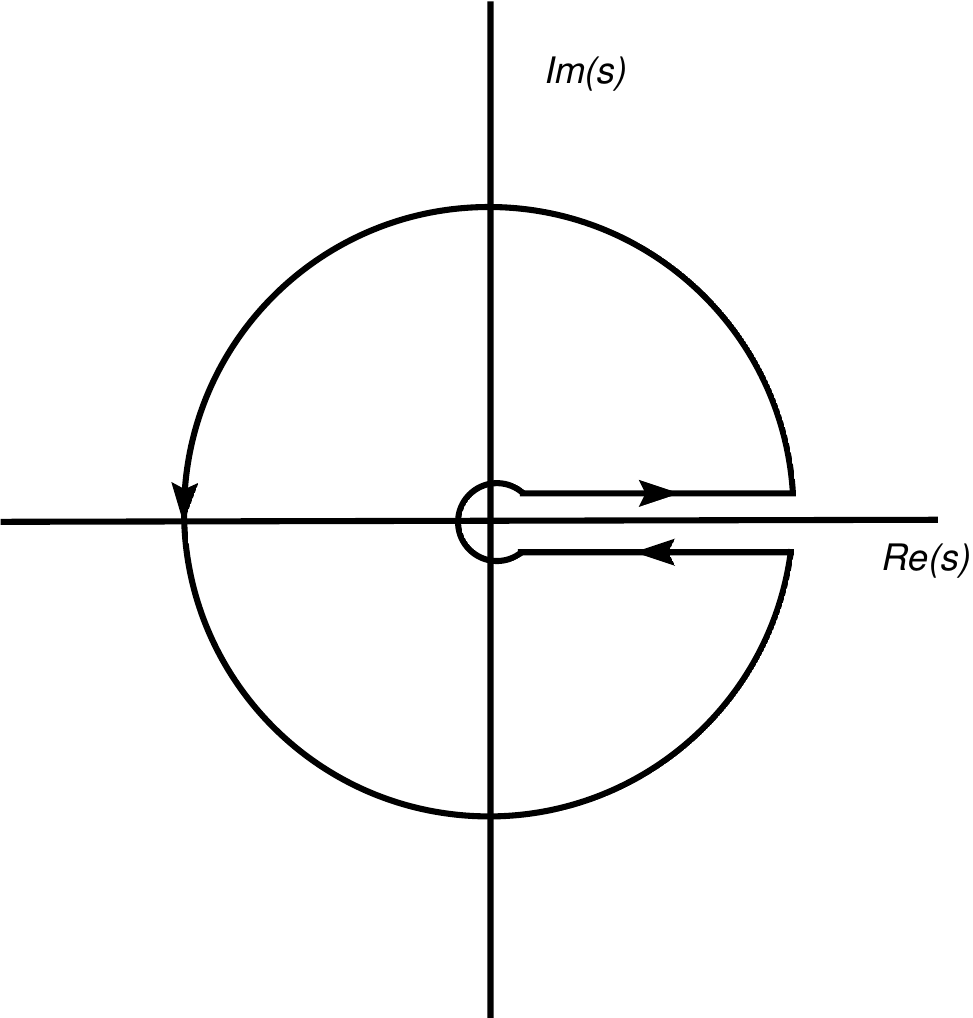}
\caption{Integration contour in the complex s-plane.}
\end{center}
\end{figure}

Cauchy's theorem applied to current correlators relates QCD information on the circle to hadronic physics on the real axis (quark-hadron duality). For the determination of the light-quark masses, $m_{u,d}$ , the appropriate correlator is that involving the axial-vector current divergences

\begin{equation}
\psi _{5}(s\equiv-q^{2})\,=\,i\int \,d^4 x\,e^{iqx}<0
|\,T(j_{5}(x)\,j_{5}(0))\,|0 >\;, \label{psi5q2}
\end{equation}

where

\begin{equation}
j_{5}(x)\, \equiv \partial ^\mu A_\mu(x)\, = \,(m_{d}+m_{u}) :\overline{d}(x)\,i\,\gamma _{5}\,u(x): \,. \label{j5QCD}
\end{equation}

Cauchy's theorem for this correlation function  becomes

\begin{equation}
\frac{1}{2\,\pi\,i} \oint_{C(|s_0|)} ds\, \psi_5(s)|_{\text{QCD}}\, P_5(s) \,+\,\int_{s_{th}}^{s_0}ds \, \frac{1}{\pi} \,{\mbox{Im}} \,\psi_5(s)|_{\text{HAD}}\, P_5(s)= \sum_i
{\mbox{R}_i} \,, \label{Cauchy2}
\end{equation}

where $P_5(s)$ is some meromorphic function, and $R_i$ the residues at the pole(s). The purpose of the function $P_5(s)$ is to quench the hadronic resonance contribution to the FESR. For the case of the pseudoscalar correlator, Eq.(\ref{psi5q2}), the hadronic spectral function involves the pion pole followed by at least two radial excitations. While the mass and width of these resonances is known, this information is hardly enough to reconstruct the hadronic  spectral function. Non-resonant background, inelasticity, resonance interference, etc. are realistically impossible to model. For these reasons the kernel $P_5(s)$ was introduced in previous quark-mass determinations  \cite{Kernels1}-\cite{Kernels3} in order to quench the contribution of the resonance region. The choice of $P_5(s)$ in the present determination will be an analytic function. Hence, there will be no residue contribution to the right-hand-side of Eq.(\ref{Cauchy2}).\\
The contour integral in Eq.(\ref{Cauchy2}) is usually performed in two ways, i.e. fixed order perturbation theory (FOPT), and contour improved perturbation theory (CIPT). In FOPT the strong coupling, $\alpha_s(s)$, is frozen on the integration contour, and the renormalization group (RG) is implemented after integration. Conversely, in CIPT the strong coupling is running and the RG improvement is used before integration. In a variety of applications either both methods give similar results, or CIPT leads to more accurate predictions. The latter will turn out to be the case in this determination.

This determination represents a substantial improvement on the previous FESR results for the up- and down- quark masses, in terms of (i) the analysis of different kernels, (ii) examining the issue of the convergence of the perturbative QCD expansion, (iii) a different implementation of the running QCD coupling, (iv) a more careful error analysis, and (v) the high numerical precision achieved in this calculation.\\
The previous determination \cite{Kernels3} performed the calculation of the quark masses in the framework of CIPT and restricted the choice of kernel to vanish at the resonance peaks, eventually preferring the kernel $P_5(s) = 1 - a_0 \, s - a_1 \, s^2$, with $a_0 = 0.897\; \mbox{GeV}^{-2}$ and $a_1 = - 0.1806 \;\mbox{GeV}^{-4}$. In the current determination, different integration kernels are considered and the calculations are done in the framework of both FOPT and CIPT.\\
Further, the issue of the convergence of the perturbative QCD expansion and its effect on the up- and down- quark masses was not addressed in \cite{Kernels3}, but will be considered in the present determination.\\
In the previous determination \cite{Kernels3}, the strong coupling was expressed in terms of the QCD scale $\Lambda_{\mbox{\scriptsize{QCD}}}$, as in $\alpha_s(s) \propto 1/\ln (s/\Lambda^2_{\mbox{\scriptsize{QCD}}})$,
a procedure that will not be followed here as it leads to unnecessary larger uncertainties. Instead, the renormalization group equation for the strong coupling will be used in order to express the coupling in terms of some well known value at a given scale, e.g. at the tau-lepton mass scale.\\
Additionally, the error analysis in \cite{Kernels3} did not include the error due the dependence of the up- and down- quark masses on the value of $s_0$; calculated the uncertainty due to the gluon condensate by gauging the effect of multiplying the gluon condensate by a factor two; and assumed, somewhat arbitrarily, a 30\% uncertainty in the hadronic sector. A more robust error analysis is given in this determination. \\

\section{Pseudoscalar current correlator in QCD}
The pseudoscalar current correlator, Eq.(\ref{psi5q2}), in QCD is given by
\begin{eqnarray}
\psi_5(q^2) &=& (\bar{m}_u + \bar{m}_d)^2 \,\Bigg\{ - q^2 \, \Pi_0(q^2) + 
{\cal{O}}(m_{u,d}^2)
  \nonumber \\ [.4cm]
&-& \frac{C_q}{-q^2} \,(\bar{m_u} + \bar{m_d}) \, \big\langle   \bar{q} \, q \big\rangle \,+\, \frac{C_4 \langle  O_4\rangle} {- q^2}  \, +\, {\cal{O}} \, \Big(\frac{1}{q^4}\Big) \Bigg\} \,, \label{psi5ud}
\end{eqnarray}

where $\bar{m}_q$ stands for the running quark mass in the MS-bar renormalization scheme. 
The perturbative QCD function, $\Pi_0(q^2)$, can be obtained from \cite{Gorishnii1}-\cite{Chet97}, whilst the $\cal{O}$$(\alpha_s^4)$ result can be found in  \cite{logmass3}. To $\cal{O}$$(\alpha_s^4)$ it is given by
\begin{equation}
\Pi_0(q^2) = \frac{1}{16 \, \pi^2} \Big[-12 + 6\, L + a_s A_1(q^2) + a_s^2  A_2(q^2)  + a_s^3  A_3(q^2) + a_s^4 A_4(q^2) \Big]\,, \label{Pi0PQCD}
\end{equation}

where $L \equiv \ln (-q^2/\mu^2)$, $a_s \equiv \alpha_s(- q^2)/\pi$, and the $A_i(q^2)$  are 

\begin{equation}
A_1(q^2)= - \frac{131}{2} + 34 \,L - 6\, L^2 \,+\, 24\, \zeta(3)\,,
\end{equation}
\begin{eqnarray}
& A_2(q^2)& = \Big(4\, n_F \,\zeta(3) -\frac{65}{4} n_F - 117 \,\zeta(3) + \frac{10801}{24}\Big) L + \Big(\frac{11}{3} n_F - 106\Big) L^2 \nonumber \\ [.4cm]
&+& \Big(- \frac{n_F}{3} + \frac{19}{2}\Big)\,L^3 + {\mbox{constants}} \,,
\end{eqnarray}

\begin{equation}
A_3(q^2) = C_1 \, L - 6 \,\Big(\frac{4781}{18} - \frac{475}{8}\, \zeta(3)\Big) \,L^2 \, + 229\, L^3 \ - \frac{221}{16} \, L^4 \,,
\end{equation}

\begin{equation}
C_1 = \frac{4748953}{864}\, - \frac{\pi^4}{6} - \frac{91519}{36}\, \zeta(3) + \frac{715}{2} \, \zeta(5),
\end{equation}

and
\begin{equation}
A_4(q^2) =  \sum_{i=1}^5 H_i\, L^i \,,
\end{equation}

with $\zeta(n)$ the Riemann zeta-function,  $n_F = 3$ in the light quark sector, and the coefficients $H_i$ involving long expressions \cite{logmass3}, numerically reducing to  $H_1 = 33 532.3$, $H_2= - 15 230.645111$, $H_3 = 3962.454926$, $H_4 = - 534.0520833$, and $H_5 = 24.17187500$. Next, the non-perturbative terms are
\begin{equation}
C_q = \frac{1}{2} +  \frac{7}{3}  \,  \, a_s \,,
\end{equation}

\begin{equation}
C_4\langle  O_4\rangle = - 
\frac{1}{8}\, a_s \, \big\langle G_{\mu\nu}\, G_{\mu\nu}\big\rangle \, \Big[ 1 + \frac{11}{2}\,a_s\Big]  \,.
\end{equation}

In FOPT one can either use the correlator $\psi_5(q^2)$, Eq.(\ref{psi5ud}), or its second derivative. However, this is not the case in CIPT, where it is far more convenient to use the second derivative, $\psi_5^{''}(q^2)$. The PQCD result for $\psi_5^{''}(q^2)$ was obtained in \cite{logmass3}, which for three flavours leads to the simplified (renormalization group improved) expression \cite{QCDSR2}, \cite{Kernels3}
\begin{equation}
\psi''_5(s)|^{\text{RGI}}_{\text{PQCD}} = - \frac{ \bar{m}_{ud}^2(s)}{4\,\pi^2}\, \frac{1}{s}\, \displaystyle\sum_{m=0}^4 K_m \, (\frac{\alpha_s(s)}{\pi})^m \label{psi2pqcd}
\end{equation}

where $s \equiv - q^2$, and
\begin{equation}
\bar{m}_{ud} (s)\equiv \frac{\bar{m}_u(s) \, + \, \bar{m}_d(s)}{2} \,,
\end{equation}

and the coefficients $K_m$ are: 
$K_0 =6$, $K_1 = 22$, $K_2 = 5071/24 - 105 \,\zeta(3)$, $K_3 = 5985291/2592 -  \pi^4/6 - 65869 \, \zeta(3) /36$ , and $K_4 = 3070.9698$.

The leading order non-perturbative terms in $\psi_5^{''}(q^2)$  are: 

\begin{equation}
\psi_5^{''}(q^2)|_{\langle G^2 \rangle} =  -  \frac{1}{4}\, \bar{m}_{ud}^2\,\frac{1}{(q^2)^3}\,\Big\langle \frac{\alpha_s}{\pi} \, G^2 \Big \rangle\,\Big(1  \, +\, \frac{11}{2}\,\alpha_s(q^2) \Big)\,, \label{gluoncond}
\end{equation}

\begin{equation}
\psi_5^{''}(q^2)|_{\langle \bar{q}\,\,q \rangle} =  \, \frac{\bar{m}_{ud}^2}{(q^2)^3}\, \bar{m}_{ud}\, \big\langle \bar{q}\,q \big\rangle\, \Big(1  \, +\, {\cal {O}}(\alpha_s) \Big)\,.
\end{equation}

Unlike the case of the correlator determining the strange-quark mass, it is safe to ignore here the quark-condensate contribution \cite{logmass2}, \cite{logmass1}.

As mentioned in the Introduction, the strong coupling is expressed in terms of a given scale $s =s^*$ where its value is known with high precision. Using the renormalization group equation for $a_s(s) \equiv \alpha_s(s)/\pi$ one can perform a Taylor expansion at some given reference scale $s = s^*$,  leading to \cite{BCK2}-\cite{Davier}
\begin{eqnarray}
&& a_s(s)\equiv \frac{\alpha_s (s)}{\pi} = a_s(s^*) + [a_s(s^*)]^2 \,(- \beta_0 \, \eta) + [a_s(s^*)]^3 (- \beta_1\,\eta + \beta_0^2\, \eta^2) 
\nonumber \\ [.3cm]
&+& [a_s(s^*)]^4 \Big(- \beta_2 \, \eta + \frac{5}{2}\, \beta_0 \, \beta_1\, \eta^2  - \beta_0^3 \, \eta^3 \Big) \nonumber \\ [.3cm]
&+& [a_s(s^*)]^5 \Big(-\beta_3 \,\eta + \frac{3}{2}\, \beta_1^2  \,\eta^2 + 3\, \beta_0 \, \beta_2\, \eta^2 - \frac{13}{3}\, \beta_0^2 \, \beta_1  \, \eta^3 + \beta_0^4 \,\eta^4 \Big)  \nonumber \\ [.3cm]
&+& [a_s(s^*)]^6 \left(- \beta_4 \,\eta + \frac{7}{2}\, \beta_0\, \beta_1\, \eta^2 + \frac{7}{2}\, \beta_0\, \beta_3\, \eta^2  - \frac{35}{6}\, \beta_0\, \beta_1^2 \,\eta^3 - 6\, \beta_0^2\, \beta_2 \,\eta^3 
 \right. \nonumber \\ [.3cm]
&+& \left. \frac{77}{12}\, \beta_0^3\, \beta_1 \,\eta^4 - \beta_0^5\, \eta^5\right) \,,\label{alpha_s*} 
\end{eqnarray}

where $\eta \equiv \ln(s/s^*)$. The beta function is
\begin{equation}
\beta(a_s) = -a_s^2 \big(\beta_0 \,+\, a_s\beta_1\, + \,a_s^2\beta_2 \, + \,a_s^3\beta_3 \, + \,a_s^4\beta_4\big) \,,\label{betaref}
\end{equation}

which is known up to $\cal{O}$$(\alpha_s^6)$ \cite{BaiChet2}. Our convention for the coefficients of the $\beta$-function for three flavours  is such that $\beta_0 =9/4$, $\beta_1 = 4$, etc. 

We use the world average of the strong coupling constant $\alpha_s(M_Z^2) = 0.1181\pm 0.0011 $ \cite{PDG}. This is run to our chosen scale $s^* = M_\tau^2$ using RunDec (version 3) to decouple over flavour thresholds \cite{Rundec}, which yields
\begin{equation}
\alpha_s(s^* \equiv \,M_\tau^2) \, = 0.3205 \pm 0.0183 \,. \label{alphatau}
\end{equation}

Similarly, by solving the renormalization group equation for $\bar{m}(s)$, the quark mass can also be expressed in terms of its value at some scale $s =s^*$ \cite{BCK2}, \cite{Mes}
\begin{eqnarray}
&&\bar{m}(s)=\bar{m}(s^*) \Bigg\{1 - a(s^*) \,\gamma_ 0 \,\eta + \frac{1}{2} \, a^2(s^*)\, \eta\, \Big[-2 \,\gamma_ 1 + \gamma_ 0\, (\beta_0 \,+ \,\gamma_ 0)\, \eta\Big]  \nonumber\\ [.3cm]
&-& \frac{1}{6} \, a^3(s^*) \,\eta \,\Big[6 \,\gamma_ 2 - 3 \,\Big(\beta_ 1\, \gamma_ 0\, + 2 \,(\beta_0 \,+\,\gamma_ 0) \, \gamma_ 1\Big)^{} \, \eta\, +\, \gamma_ 0 \,(2 \,\beta_0^2 \,+ 3 \,\beta_0 \,\gamma_ 0 \,+\, \gamma_0^2) \,\eta^2 \Big] \nonumber\\ [.3cm]
&+& \frac{1}{24}\,  a^4(s^*) \, \eta \, \Big[-24\, \gamma_3 \,+ \,12 (\beta_ 2\,
 \gamma_ 0\, +\, 2 \beta_ 1\,\gamma_ 1 \,+ \,\gamma_ 1^2 \,+ 3\, \beta_0\, \gamma_ 2 \,+ 2\, \gamma_ 0 \,\gamma_ 2)\, \eta\,  
\nonumber\\ [.3cm]
& -& \, 4\, \Big(6\, \beta_0^2\, \gamma_ 1 \,+ \, 3\, \gamma_ 0^2\, (\beta_ 1 \,+ \,\gamma_1) \,+\, \beta_0\, \gamma_ 0 \,(5 \,\beta_ 1\, + \,9\, \gamma_ 1)\Big) \,\eta^2 \,+ \,\gamma_ 0 \,(6\, \beta_0^3\, +\, 11\,\beta_0^2 \,\gamma_ 0 \,
\nonumber\\ [.3cm]
&+& 6\, \beta_0\, \gamma_ 0^2\, +\, \gamma_ 0^3) \,\eta^3 \Big] \nonumber\\ [.3cm]
&+& \dfrac{1}{120}\, a^5(s^*)\, \eta \,\Big[-120\, \gamma_ 4 \,+\,  \dfrac{1}{\beta_0} 60\, \Big(-7 \,\beta_ 1 \,\beta_ 2\, \gamma_0 \,+\, 4 \,\beta_0^2\, \gamma_ 3\, +\, \beta_0\,(7 \,\beta_ 1 \,\gamma_ 0 \,+ \,\beta_ 3 \,\gamma_ 0\,
\nonumber\\ [.3cm]
&+& 2\, \beta_ 2\, \gamma_ 1 \,+ \,3\,\beta_ 1 \,\gamma_ 2\, +\,  2 \gamma_ 1 \,\gamma_ 2 
\, +\, 2\,\gamma_ 0 \,\gamma_ 3)\Big)\, \eta \,- \,20\,\Big(3 \beta_ 1^2 \,\gamma_ 0 \,+\, \beta_ 1\, (14 \, \beta_0 \,+\,
9 \,\gamma_ 0)\, \gamma_ 1 \,
\nonumber\\ [.3cm]
&+& 3\, (2 \beta_0\, +\, \gamma_ 0) (\beta_ 2\, \gamma_ 0\, + \,\gamma_ 1^2\, + \,
2\,\beta_0\, \gamma_ 2 \,+\, \gamma_ 0\, \gamma_ 2)\Big) \,\eta^2
\,+\,10 \Big(12 \,\beta_0^3\, \gamma_ 1 \,+ \,\gamma_ 0^3 (3 \, \beta_ 1 \,+ \,
2 \,\gamma_ 1) 
 \nonumber\\ [.3cm]
&+ & \beta_0\, \gamma_ 0^2\, (13\, \beta_ 1 \,+\, 12\, \gamma_ 1) \,+ \,
\beta_0^2\, \gamma_ 0\, (13 \,\beta_ 1\, + \,22\,\gamma_ 1)\Big) \,\eta^3 
\,-\, \gamma_0\, \Big(24\, \beta_0^4 \,+\, 50\, \beta_0^3\, \gamma_ 0 
 \nonumber\\ [.3cm]
&+ & 35\, \beta_0^2\,\gamma_0^2 \,+\, 10\, \beta_0 \,\gamma_0^3\, + \,\gamma_0^4\Big) \,\eta^4 \Big] 
\,+\, \mathcal{O}(a^6(s^*))\Bigg\} \,, \label{m_qs*}
\end{eqnarray}

where the $\gamma(a_s)$ function is \cite{BaiChet3}
\begin{equation}
\gamma(a_s) = -a_s \big(\gamma_0 \,+\, a_s\gamma_1 \,+\, a_s^2\gamma_2\,+\, a_s^3\gamma_3 \,+\, a_s^4\gamma_4\big) \label{gammadef}
\end{equation}

with the convention such that e.g. $\gamma_0 = 1$, $\gamma_1 = 91/24$, etc., for three flavours.

\section{Hadronic pseudoscalar current correlator}
In the hadronic sector, the spectral function of the current correlator $\psi_5(q^2)$, Eq.(\ref{psi5q2}), involves the pion pole
followed by the three-pion resonance contribution

\begin{equation}
\frac{1}{\pi} {\mbox{Im}} \, \psi_5|_  {\text{HAD}}(s)= 2\, f_\pi^2 \, M_\pi^4 \, \delta(s - M_\pi^2) \, + \,\frac{1}{\pi} \, {\mbox{Im}}\,
\psi_5|_\text{RES}(s) \,
\end{equation}

where $f_\pi = (92.07\, \pm \, 1.20) \,$ MeV \cite{PDG},  $M_\pi = (134.9770\, \pm \, 0.0005)$ MeV \cite{PDG}, and the three-pion resonance contribution is due to the $\pi(1300)$ followed by the  $\pi(1800)$ \cite{PDG}. In the chiral limit the threshold behaviour  of the three-pion state, first obtained in \cite{PagelsZepeda}, is 

\begin{equation}
	\frac{1}{\pi}\; {\mbox{Im}} \;\psi_5(s)|_{\pi\pi\pi} \; = \theta(s)  \;\frac{1}{3}\;\frac{M_\pi^4}{f_\pi^2} \frac{1}{2^8\,\pi^4}\, s\; \;. \label{PagelsZ}
\end{equation}

Beyond the chiral limit the threshold behaviour, first obtained in \cite{CADdeR} and later corrected for misprints in \cite{Prades}, is given by 

\begin{equation}
	\frac{1}{\pi}\; {\mbox{Im}} \;\psi_5(s)|_{\pi\pi\pi} \; = \theta\big(s- 9\, M_\pi^2\big)  \;\frac{1}{9}\;\frac{M_\pi^4}{f_\pi^2} \frac{1}{2^8\,\pi^4}\, I_{PS}(s)\; \;.
\end{equation}

where the phase-space integral $I_{PS}(s)$ is 

\begin{eqnarray}
	I_{PS}(s)\;&=&\; \int_{4 M_\pi^2}^{(\sqrt{s} - M_\pi)^2}\; du \;\sqrt{1 - \frac{4M_\pi^2}{u}} \; \;\lambda^{1/2}(1,u/s,M_\pi^2/s)\;
	\Biggl\{5 + \frac{1}{2} \;\frac{1}{(s-M_\pi^2)^2}
	\Biggr. \nonumber \\ [.3cm]
	&\times& \left.   \Bigl[(s - 3 u + 3 M_\pi^2)^2 
	+  3\; \lambda(s,u,M_\pi^2) \;\left(1 - \frac{4\, M_\pi^2}{u}\right) + 20 \;M_\pi^4 \Bigr]  \Biggr. \right.\nonumber \\ [.3cm]
	&+& \Biggl. \frac{1}{(s - M_\pi^2)}\; \Bigl[ 3 (u - M_\pi^2) - s + 9 M_\pi^2 \Bigr] \Biggr\} \;, \label{IPS}
\end{eqnarray}

where
\begin{equation}
	\lambda(1,u/s,M_\pi^2/s) \equiv \Bigl[1 - \frac{\left(\sqrt{u} + M_\pi\right)^2}{s}\Bigr]
	\; \Bigl[1 - \frac{\left(\sqrt{u} - M_\pi\right)^2}{s}\Bigr]\;,
\end{equation}

\begin{equation}
	\lambda(s,u,M_\pi^2) \equiv \Bigl[s - \left(\sqrt{u} + M_\pi\right)^2\Bigr]
	\; \Bigl[s - \left(\sqrt{u} - M_\pi\right)^2\Bigr]
	\; ,
\end{equation}

which in  the chiral limit it reduces to $I_{PS} = 3\, s$.

This threshold expression normalizes the hadronic resonance spectral function, modelled as a combination of Breit-Wigner forms $BW_i(s)$

\begin{equation}
	\frac{1}{\pi}\; {\mbox{Im}} \;\psi_5(s)|_\text{RES} \; = {\mbox{Im}} \;\psi_5(s)|_{\pi\pi\pi} \,\frac{[BW_1(s) + \kappa \;BW_2(s)]}{(1+\kappa)} \;, \label{ImPi5}
\end{equation}

where $BW_1(s_{th}) = BW_2(s_{th}) = 1$, with

\begin{equation}
	BW_i(s) = \frac{(M_i^2 - s_{th})^2 + M_i^2 \,\Gamma_i^2}{(s - M_i^2)^2 + M_i^2 \Gamma_i^2}\;\;\;\;(i=1,2)\;, \label{BWi}
\end{equation} 

and $\kappa$ is a free parameter controlling the relative weight of the resonances. 
The value $\kappa = 0.1$ results in a smaller contribution of the second resonance compared to the first, and it will be used in the sequel. The widths of these radial excitations of the pion are affected by large uncertainties \cite{PDG}. For the first resonance, $\pi$ (1300) we shall use the determination from the two-photon process $\gamma \, \gamma \, \rightarrow \, \pi^+\, \pi^-\, \pi^0$, as it is the most reliable \cite{3pionres}. The width is $\Gamma_1 =(260 \pm 36)$ MeV.  The second resonance is the $\pi(1800)$, with a width  $\Gamma_2 =(208 \pm 12)$ MeV \cite{PDG}.\\

\begin{figure}[ht]
\begin{center}
\includegraphics[height=3.0in, width=4.6in]
{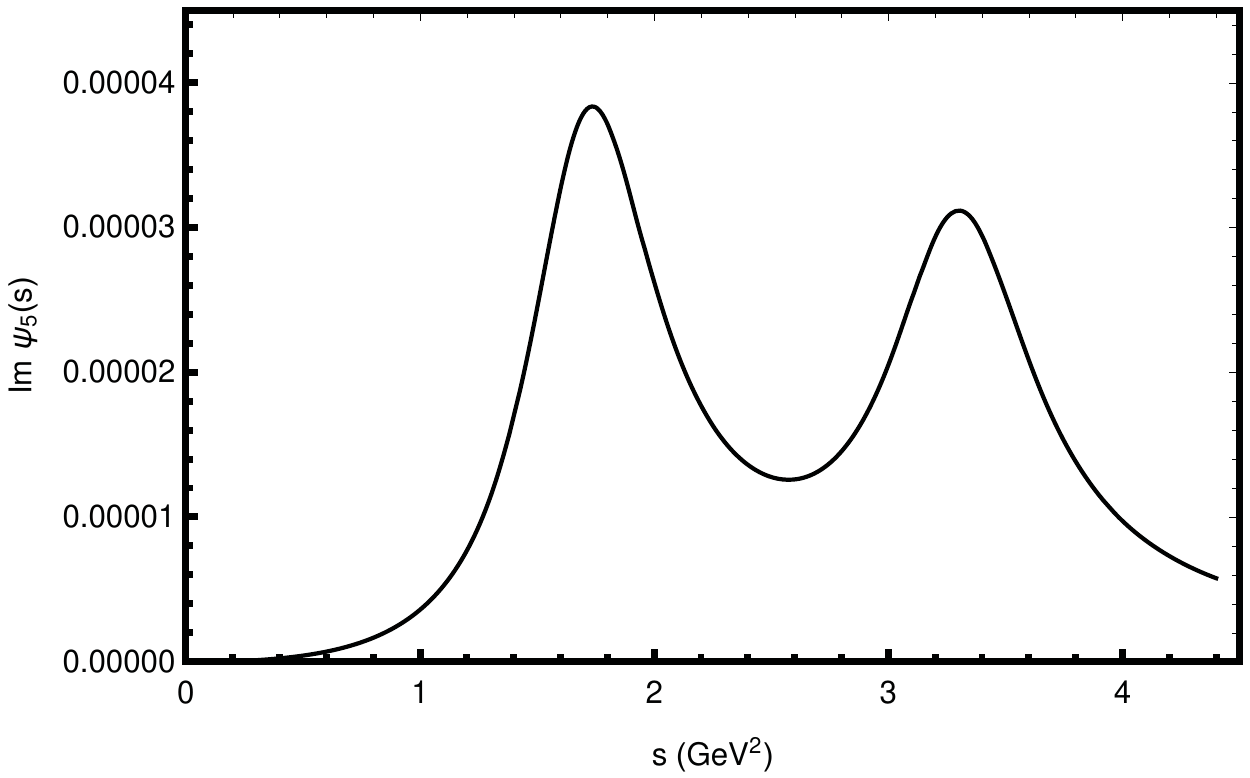}
\caption{Hadronic spectral function in the resonance region, Eqs.(\ref{ImPi5})-(\ref{BWi}) with $\kappa = 0.1$, and  involving two radial excitations of the pion, $\pi(1300)$ and $\pi(1800)$.}
\end{center}
\end{figure}

\section{QCD sum rules and results}
\begin{figure}[ht] 
\begin{center}
\includegraphics[height=2.9in, width=4.6in]
{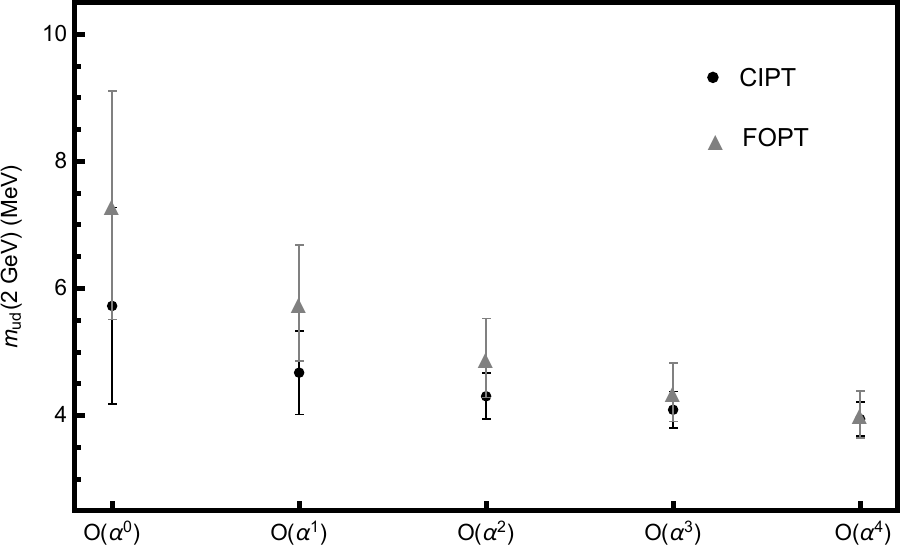}
\caption{The quark mass $\bar{m}_{ud}$(2 GeV) plotted by successively including higher order terms in the perturbative expansion of the correlator function before integration, in both CIPT and FOPT.}\label{convergencegraph}
\end{center}
\end{figure}

The starting point is the analysis of the convergence of the correlator function's PQCD expansion using FOPT. In FOPT, the strong coupling is fixed for a given radius $s_0$ in the complex s-plane. After the contour integration is performed one finds a series in terms of $\alpha_s(s)$, Eq.(\ref{seriespade}), the convergence of which can be analysed. A remark must be made that this is not the case in CIPT where the strong coupling is running, i.e. its value must be found by solving the relevant renormalization group equation at each point along the contour. As such, the contour integration in CIPT must be performed numerically and no symbolic series in terms of $\alpha_s(s)$ can be found. Hence, the convergence of PQCD expansion of the correlator function can not be directly analysed in CIPT. This does not, however, preclude one from analyzing the convergence of the quark mass in both FOPT and CIPT, by successively including higher order terms in the perturbative expansion of the correlator function before integration, which is addressed later in this paper (Fig. \ref{convergencegraph}). 

The quark mass, $\bar{m}_{ud}(s_0)$, is determined in FOPT from the FESR, Eq.(\ref{Cauchy2}), as

\begin{equation}
(\bar{m}_u + \bar{m}_{d})^2 = \frac{\delta_5(s_0)|_\text{HAD}}
{\delta_5(s_0)|_\text{QCD}} \,, \label{mq^2}
\end{equation}

\begin{equation}
\delta_5(s_0)|_{\text{HAD}} \, =  \int_{s_{th}}^{s_{0}} ds \, \frac{1}{\pi} \, \text{Im}\, \psi_5(s)|_{\text{HAD}} \, P_5(s) \,,
\end{equation}

\begin{equation}
\delta_5(s_0)|_\text{QCD} =  -\frac{1}{2\pi i} \, \oint_{C(|s_0|)} ds\, \hat\psi_5(s)|_{\mbox{\scriptsize{QCD}}} \, P_5(s) \,,
\end{equation}

where $\hat \psi_5(s)|_{\mbox{\scriptsize{QCD}}} $ stands for the correlator, Eq.(\ref{psi5ud}), with the overall quark-mass squared factor removed, and  $P_5(s)$ is an analytic integration kernel designed to quench the hadronic contribution to the sum rule. Notice that the dimension, $d$,  of $\delta_5(s_0)|_{\text{HAD}}$ is $d=6$, while that of $\delta_5(s_0)|_{\text{QCD}}$ is $d=4$. Regarding $P_5(s)$, several functional forms for the hadronic quenching integration kernel were considered, with the optimal being

\begin{equation}
P_5(s) = (s - c) (s - s_0) \,, \label{Delta_5}
\end{equation}

where $c = 2.4 \, {\mbox{GeV}}^{-2}$ lies halfway between the two resonances. Several criteria were used in choosing the integration kernel Eq.(\ref{Delta_5}). For instance, the kernel should not bring in higher dimensional condensates, as their values are poorly known. This constrains substantially the powers of $s$. Next, the relative contribution of the second resonance should not exceed that of the first one. The kernel should quench the hadronic resonance contribution at $s = s_0$, as well as in the region between the two resonances. The kernel Eq.(\ref{Delta_5}) also leads to the most stable result for the quark masses in the wide region $s_0 \simeq (1.5 - 4.0) \mbox{GeV}^2$.

Substituting the PQCD result, as given in Eqs.(\ref{psi5ud})-(\ref{Pi0PQCD}), at a typical scale of $s_0 =3.3 \, \text{GeV}^2$, leads to
\begin{equation}\label{seriespade}
[\delta_5(s_0)|_{\text{PQCD}}]^{-1/2} = 2.42 \, (1 \, + \,  2.68\, \alpha_s \, + \, 8.63\, \alpha_s^2 \, + \, 25.77 \,\alpha_s^3 \, + \, 71.63 \,\alpha_s^4)^{-1/2} \,,
\end{equation}

in units of $\mbox{GeV}^{-2}$, and  $\alpha_s \equiv \alpha_s(s_0)$. Using Eqs.(\ref{alpha_s*})-(\ref{alphatau}) to obtain $\alpha_s(s_0)$ shows that all terms beyond the leading order are roughly of the same size

\begin{equation}\label{seriespade2}
[\delta_5(s_0)|_{\text{PQCD}}]^{-1/2} = 2.42 \, (1 \, + \,  0.85 \, + 0.86 + \, 0.82 \, + \, 0.72)^{-1/2}\,.
\end{equation}

Since the quark mass actually depends on the square-root of $\delta_5$, the relevant power series expansion is instead
\begin{equation}\label{padeseries}
[\delta_5(s_0)|_{\text{PQCD}}]^{-1/2}=  2.42 \, (1 \, - \,  1.34\, \alpha_s \, - \, 1.62\, \alpha_s^2 \, - \, 1.55 \,\alpha_s^3 \, -\, 0.11  \,\alpha_s^4)\,.
\end{equation} 

Substituting in $\alpha_s(s_0)$ (found from Eqs.(\ref{alpha_s*})-(\ref{alphatau})), Eqs.(\ref{padeseries}) becomes

\begin{equation}\label{delta5QCDconv}
[\delta_5(s_0)|_{\text{PQCD}}]^{-1/2} = 2.42 \, (1 \, - \,  0.42 \, - 0.16 \, - \, 0.05 \, - \, 0.001) \,, 
\end{equation}

which shows a much improved convergence. Interestingly, the expansion, Eq.(\ref{padeseries}), is an example of a Pad\'{e} approximant; in this case a [4/0] approximant.  As a consequence of this we have tried other types of Pad\'{e} approximants, but this simple one provides the optimal expansion in this application. While this Pad\'{e} improvement is unquestionably a positive feature, there remain other unwelcome issues with FOPT. These include a large negative impact on the results for $\bar{m}_{ud}$ from (i) the dependence of the results on the value of $s_0$, (ii) the estimate of the unknown six-loop contribution, and (iii) the uncertainties in $\alpha_s$ when using Pad\'{e} approximants. These issues are under much better control in CIPT, which is described next.  

In the framework of  CIPT the QCD sum rule is given by \cite{QCDSR2}

\begin{eqnarray}
 &-& \frac{1}{2 \pi i} \oint_{C(|s_0|)}	ds \, \psi^{''}_5(s)|_\text{QCD} \; \big[F(s) - F(s_0)  \big] 
 \nonumber\\ [.3cm]
&=& 2 \,f_\pi^2 \, M_\pi^4 \, P_5(M_\pi^2) +
 \frac{1}{\pi}\, \int_{s_{th}}^{s_0} ds \, {\mbox{Im}} \, \psi_5(s)|_{\text{RES}} \,\, P_5(s)\,, \label{FESRlhs}
 \end{eqnarray}
where $F(s)$ depends on the explicit form of the kernel $P_5(s)$. The function $F(s)$ corresponding to this integration kernel is given by

\begin{equation}
F(s) = \frac{1}{12}\, s^4 - \frac{1}{6} (c + s_0)\, s^3 + \frac{1}{2}\, c \, s_0 \, s^2 + \Big(\frac{s_0^3}{6} - \frac{1}{2}\, c\, s_0^2 \Big)\, s \,, \label{Fs}
\end{equation}

and $F(s_0)$ becomes
\begin{equation}
F(s_0) = \frac{s_0^3}{12}\, (- 2 \,c + s_0)\,. \label{Fs0}
\end{equation}

After substituting Eqs.(\ref{psi2pqcd})  in Eq.(\ref{FESRlhs}) the left-hand-side of the FESR, Eq.(\ref{FESRlhs}),  becomes (after renormalization group improvement)
\begin{equation}
\delta_5(s_0)|^{\mbox{\scriptsize{RGI}}}_{\mbox{\scriptsize{{PQCD}}}} = \frac{\bar{m}^2_{ud}}{16\,\pi^2} \,\sum_{n=0}^{4} K_n \,\frac{1}{2 \pi i} \oint_{C(|s_0|)} \frac{ds}{s}\,\Big[F(s) - F(s_0) \Big]\, \Big(\frac{\bar{\alpha}_s(s)}{\pi} \Big) ^n\,, \label{delta5pqcd}
\end{equation}

with the coefficients $K_n$ defined in Eq.(\ref{psi2pqcd}). After substituting Eqs. (\ref{Fs}) and (\ref{Fs0}) into Eq.(\ref{delta5pqcd}), there are two types of integrals involved, to be computed numerically,
\begin{equation}
I^a_{NM}(s_0) \equiv \frac{1}{2\,\pi i} \, \oint \frac{ds}{s} \, s^N \left( \frac{\bar{\alpha}_s}{\pi} \right)^M
\end{equation}

and 
\begin{equation}
I^b_{M}(s_0) \equiv \frac{1}{2\,\pi i} \, \oint \frac{ds}{s} \,\left(\frac{\bar{\alpha}_s}{\pi} \right)^M \,,
\end{equation}

where $N$ and $M$ are positive integers.

Finally, the running of the quark mass must be taken into account. This is achieved by starting from the RG equation for the mass

\begin{equation}
\frac{dm}{m} = \frac{\gamma(\alpha_s)}{\beta(\alpha_s)} \,d \alpha_s \, \label{massrge}
\end{equation}

where $\gamma(\alpha_s)$ and $\beta(\alpha_s)$ were defined in Eqs.(\ref{betaref}), (\ref{gammadef}).

After the change of variables $d \alpha_s(x) = i \beta(\alpha_s)\, dx$ in Eq.(\ref{massrge}), followed by integration,  Eq.(\ref{massrge}) becomes
\begin{equation}
m(x) = m(x_0) \, exp\, \Big[ i \, \int_{x_0}^{x} \gamma[\alpha_s(x')] \, dx' \Big]\,,
\end{equation}

and the running quark mass entering the FESR is given by

\begin{equation}
\bar{m}_{ud}(x) = \bar{m}_{ud}(s_0) \, exp \Bigg[- i \, \int_0^x  dx' \, \sum_J \gamma_J \Big[a_s(x') \Big]^J \Bigg]\,,
\end{equation}

such that the FESR determines $\bar{m}_{ud}(s_0)$. The initial value of the strong coupling is obtained from Eqs.(\ref{alpha_s*})-(\ref{alphatau}).

In the non-perturbative sector we use the value of the gluon condensate in Eq.(\ref{gluoncond}) from a recent precision determination \cite{GG2}  (earlier determinations are discussed in detail in \cite{QCDSR2}) 

\begin{equation}
\Big\langle  \frac{\alpha_s}{\pi} G^2 \Big\rangle = (0.037 \,\pm\, 0.015) \, {\mbox{GeV}}^4 \;. \label{GGJpsi}
\end{equation}

In the hadronic sector the spectral function is parametrised as in Eqs.(\ref{ImPi5})-(\ref{BWi}). The parameter $\kappa$ can be varied in the  wide range $\kappa = 0.1 -0.2$,  subject to the requirement that the first resonance should be leading. Such a variation produces only a 1\%  change in $\bar{m}_{ud}$.

In calculating $\bar{m}_{ud}$, the results from FOPT and CIPT are in agreement (see supplementary calculations for more detail). However based on two central criteria - stability and convergence - CIPT is the preferred framework for determining the $\bar{m}_{ud}$. The results are given below.\\ 

\begin{figure}[ht]
\begin{center}
\includegraphics[height=3.0in, width=4.6in]
{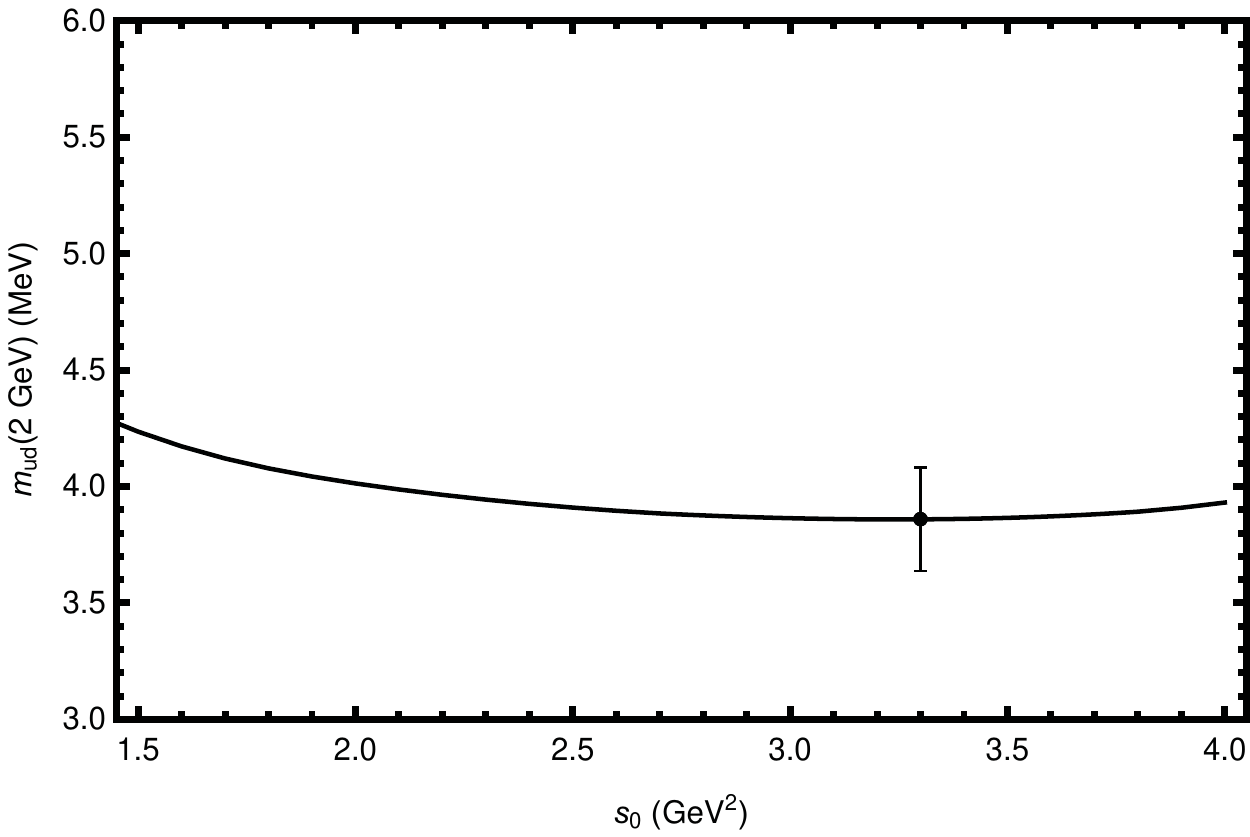}
\caption{The quark mass $\bar{m}_{ud}$(2 GeV) as a function of $s_0$ in CIPT from the FESR, Eq.(\ref{FESRlhs}).} \label{stabilitygraph}
\end{center}
\end{figure}

The result for $\bar{m}_{ud}$ as a function of $s_0$ from the sum rule, Eq.(\ref{FESRlhs}), in CIPT  is shown in Fig. \ref{stabilitygraph}. Potential, in principle unknown \cite{GonzalezViolation}-\cite{PichViolation}, duality violations are expected to be quenched above $s_0 \approx 3.0 \, \text{GeV}^2$, the region where the result is obtained.\\

\begin{table}[ht]
	\begin{tabular}{cccccccccc}
		\cline{1-9}
		\noalign{\smallskip}
		& $\,\bar{m}_{ud}(2\,\text{GeV})\,\,\,$ & $\Delta_{\alpha_s}\,\,$ & $\Delta_{ \langle G^2\rangle}\,\,$ &	$\Delta_{ s_0}\,\,$ &		$\Delta_{ \text{HAD}}\,\,$ & $\Delta_{\text{6-loop}}\,\,$ & $\Delta_{\text{T}}\,$ \\   
		& (MeV)&&&&&&&\\
		\hline 
		\noalign{\smallskip}
		CIPT &3.946 & 0.207	& 0.052 & 0.017	& 0.084 & 0.132 & 0.265\\
		\hline
	\end{tabular}
	\caption{Results for the various uncertainties from CIPT, together with the total uncertainty added in quadrature, $\Delta_T$.}
\end{table}

The error bar in Figure \ref{stabilitygraph} is the total uncertainty due to the various sources shown in Table 1.
These are (i) the uncertainty in the strong coupling, $\alpha_s$, Eq.(\ref{alphatau}), (ii) the uncertainty in the value of the gluon condensate, Eq.(\ref{GGJpsi}), (iii) the range $s_0 = (1.5 - 4.0) \, \mbox{GeV}^2$, (iv) the uncertainty in the resonance widths and the parameter $\kappa$ in the hadronic spectral function, and (v) the assumption that the unknown PQCD six-loop contribution is equal to the five-loop one.  This leads to

\begin{equation}
\bar{m}_{ud}|_{\mbox{\scriptsize{CIPT}}} (2 \,\mbox{GeV} ) = (3.9\, \pm 0.3)\,{\mbox{MeV}} \,,
\end{equation}

to be compared with the PDG value \cite{PDG} $\bar{m}_{ud}|_{\mbox{\scriptsize{PDG}}} (2 \,\mbox{GeV} ) = (3.5\,\pm 0.6)  \,{\mbox{MeV}}$, and
the FLAG Collaboration result \cite{FLAG}  $\bar{m}_{ud}|_{\mbox{\scriptsize{FLAG}}} (2 \,\mbox{GeV} ) = (3.373\, \pm 0.080)  \,{\mbox{MeV}}$. In order to disentangle the individual mass values one requires as external input the quark mass ratio $m_u /m_d $. \\

Using the recent PDG value \cite{PDG}

\begin{equation}
\frac{m_u}{m_d} = 0.48 \pm 0.08\,,
\end{equation}

results in
\begin{equation}
\bar{m}_u (2\, \mbox{GeV}) = (2.6 \, \pm \, 0.4) \, {\mbox{MeV}} \,, \label{mufinal}
\end{equation} 

\begin{equation}
\bar{m}_d (2\, \mbox{GeV}) = (5.3 \, \pm \, 0.4) \, {\mbox{MeV}} \,,
\label{mdfinal}
\end{equation}

to be compared with the PDG values \cite{PDG}: $\bar{m}_u \,(2\, \mbox{GeV}) = (2.2 \,\pm 0.5)\, \,{\mbox{MeV}}$, and $\bar{m}_d \,(2\, \mbox{GeV}) = (4.7 \,\pm 0.5)$\,{\mbox{MeV}}; and with the FLAG Collaboration results \cite{FLAG} $\bar{m}_u \,(2\, \mbox{GeV}) = (2.16 \, \pm 0.11) \, \,{\mbox{MeV}}$, and $\bar{m}_d \,(2\, \mbox{GeV}) = (4.68 \, \pm 0.16) \, \,{\mbox{MeV}}$. \\

{\bf Acknowledgements:} This work was supported in part by the Alexander von Humboldt Foundation (Germany), under the Research Group Linkage Programme, and by the University of Cape Town (South Africa). The authors wish to thank Hubert Spiesberger for discussions.\\

{\bf Notice:} The Mathematica code used in the numerical evaluations  is attached as a supplementary resource.

\end{document}